\documentclass[twocolumn,showpacs,preprintnumbers,amsmath,amssymb]{revtex4}

\usepackage{amsmath}
\usepackage{graphicx}
\usepackage{dcolumn}
\usepackage{bm}

\begin{document}

\title{Creep failure in a threshold activated dynamics: Role of temperature during a sub-critical loading}
\author{Subhadeep Roy}
\email{subhadeep.roy@ntnu.no} 
\affiliation{PoreLab, Department of Physics, Norwegian University of Science and Technology, NO--7491 Trondheim, Norway}

\author{Takahiro Hatano}
\email{hatano@ess.sci.osaka-u.ac.jp}
\affiliation{Department of Earth and Space Science, Osaka University, 560-0043 Osaka, Japan}

\date{\today {}}
\begin{abstract}
Creep is time-dependent deformation of solids at relatively low stresses, leading to the breakdown with time. Here we propose a simple model for creep failure of disordered solids, in which temperature and stress are controllable. Despite its simplicity, this model can reproduce most experimental observations. Time dependence of the strain rate is well fitted with power laws resembling the Omori-Utsu and the inverse Omori laws in the primary and the tertiary creep regimes, respectively. Distribution of the creep lifetime obeys the log-normal distribution, and the average creep lifetime decays in a scale-free manner with the increasing stress. The above results are in good agreement with experiments. Additionally, the mean avalanche size as a function of temperature exhibits a series of jumps, and finite size scaling implies the existence of phase transitions. 
\end{abstract}
\maketitle


\section{Introduction}
Failure point and the failure processes of disordered solids and composite materials depend on temperature, pressure, and driving conditions such as applied stress or strain rate \cite{Lawn}. Importantly, materials can deform and break with time even if the applied stress is below the critical value. This phenomenon is called as creep failure or creep rupture. Study of creep failure is essential in many contexts such as constructions, in which the building blocks are subject to the load for a considerable duration. The time elapsed until creep failure is known as the creep lifetime. From the practical point of view, estimate of creep lifetime leads to a direct forecast of the catastrophic failure event and therefore important in materials science. The creep lifetime depends on many ingredients, among which the applied stress and the temperature have prominent effects \cite{Wong}.

Phenomenology of creep includes several power laws. Suppose that the stress is applied to a specimen at $t = 0$ and kept constant thereafter. Then the specimen starts to deform, but the strain rate decays with time as $t^{-p}$ \cite{Andrade}. After a certain lapse of time, deformation accelerates toward breakdown with the strain rate increasing as $(\tau_c-t)^{-p'}$, where $\tau_c$ denotes the creep lifetime of the specimen. The two exponents, $p$ and $p'$, are not generally the same, but typically range from $0.6$ to $1.0$ \cite{Andrade,Nechad2005PRL,Nechad2005JMPS,Leocmach,Miguel}. The latter power law describing acceleration toward breakdown has been of interest from practical point of view, since the time of catastrophic breakdown could be predicted by fitting the real-time data to the above formula \cite{Main1999,Kilburn,Vasseur}. Interestingly, similar power laws are also known for earthquakes \cite{Omori,Utsu,Jones}, which involve much larger scales than the experimental specimens. In this case, the creep lifetime may correspond to the time of earthquake. Therefore, to understand the physics behind these power laws may lead us to some novel principles that are common to fracture phenomena across different scales, as well as to develop useful tools for forecasting catastrophic failure events.

Since the physical mechanisms behind creep may be the accumulation of damage and plastic strain in a system, models for creep should take such aging processes into account. Structural randomness and heterogeneity in solids should be also considered appropriately. However, the fracture of heterogeneous solids is yet hard to handle with the elasticity theory for continuum, and therefore there have been some alternative approaches. A pioneering work by Main \cite{Main2000} is based on the subcritical crack growth dynamics and the phenomenological Voight's model for precursory strain \cite{Voight1989}. Assuming a heterogeneous feedback system, the model can reproduce both the power laws with the exponents directly related to those assumed at the microcrack level. Another class of approaches makes use of a simple and intuitive model for disordered solids 
 \cite{Nechad2005JMPS,Ciliberto,Scorretti,Politi,Saichev,Hidalgo2001,Hidalgo2002,Kun2003,Pradhan2003,Danku}, which is  known as the fiber bundle model \cite{Pierce,Daniels,Herrmann,Chak1,Chak2,Fiber1,Fiber2}. Since temperature plays an important role in creep rupture, thermal fluctuations are explicitly incorporated in these models in the form of noise or probabilistic time evolution. They appear to be successful in explaining the power law behaviors of the time-dependent strain rate as well as the temperature and the stress (or strain) dependences of the creep lifetime \cite{Ciliberto,Scorretti,Politi,Saichev}. Instead of introducing stochasticity, Hidalgo et al. introduce the time evolution equation for the strain of each constituent (i.e., fiber) based on the Kelvin-Voigt rheology \cite{Hidalgo2002,Kun2003}, and obtain the temperature and the stress dependences of creep lifetime that are somewhat different from those in the above-mentioned stochastic models \cite{Ciliberto,Scorretti,Politi,Saichev}. Danku and Kun consider a damage accumulation process in each fiber by introducing the evolution equation for the damage variable \cite{Danku}. Their model reproduces the inverse Omori law with exponent $p'=1$, although it increases up to 5 if the smaller events are disregarded.

Contrastingly, much simpler models can exhibit creeplike behaviors even in the absence of any thermal fluctuations, damage variables, or rheological constitutive laws \cite{Pradhan2001,Royarxiv,royhatano1}. In particular, the present authors derived both the power laws for creep only by introducing the time evolution in a simple fiber bundle model \cite{royhatano1}. However, the obtained exponents ($p=1.8$ and $p'=2$) are too large to be comparable to any experimental values. 

Here we propose a slightly modified model, in which the effect of temperature is included. Probabilistic algorithm is adopted for the failure criterion of fibers, bringing thermal fluctuations in the model in addition to the time-independent (i.e., quenched) randomness. The interplay between these two kinds of randomness yields nontrivial time-dependent behaviors. We have numerically studied the model focusing on the effects of temperature and applied stress on the creep behaviors: statistics of the creep lifetime including the distribution function, the time evolution of strain rate, and the avalanche statistics. In the next section, we have provided a detailed description of the conventional fiber bundle model along with the modifications we have made for the present work. This is followed by the numerical results together with some brief comments and discussions on the future scope as a continuation of the present observations.


\section{Description of the model}\label{Sec2}
After its introduction by Pierce in 1926 \cite{Pierce}, the fiber bundle model has been proven to be a useful and yet arguably the simplest model to understand the failure process in disordered solids \cite{Daniels,Chak2,Fiber1,Fiber2}. The fiber bundle model consists of $L$ vertical fibers in between two parallel bars. A load $F$ is applied to the bars to create a stress $\sigma=F/L$ per fiber. Each fiber has an individual strength chosen from a threshold distribution randomly. This heterogeneity in the strength may be regarded as a quenched randomness in the model, and the dispersion of threshold distribution measures the strength of disorder in the model. When the applied stress exceeds a threshold value, the corresponding fiber breaks irreversibly, and the load borne by that fiber is redistributed within the model; either among all the surviving fibers (mean field model) or among the surviving nearest neighbors only (local stress concentration). Due to such redistribution, the local stress values of some fibers increase and that can lead to further breaking and redistribution. This is an avalanche. After a certain number of avalanches, the model breaks completely or relaxes to a stable state with nonzero surviving fibers. In the latter case, the external load $F$ needs to increase to break the next weakest fiber leading to further avalanches. This process continues until all the fibers are broken. The applied load just before the global failure is referred to as the critical load $F_c$.

Fiber bundle model has some variations to include thermal effects in fracture \cite{Ciliberto,Scorretti,Politi,Coleman,Roux,Pradhan2003,Yoshioka10,Yoshioka12}. Here we also adopt a probabilistic rule that is dominated by temperature. Let us assume that any fiber breaks with a certain probability that depends on the  temperature $T$. In this study, the probability of the $i$th fiber to break at time $t$ is given by
\begin{equation}\label{Eq.1}
P_r(t,i) = P'\exp\left[\frac{{\sigma(t,i)}-\sigma_{\rm th}(t,i)}{T}\right].
\end{equation}
Here $\sigma(t,i)$ and $\sigma_{\rm th}(t,i)$ are, respectively, the local stress and the stress threshold of $i$th fiber at time $t$, and $P^{\prime}$ is a constant chosen to be unity hereafter. Note that any fibers can rupture even if the local stress is less than its threshold value. If $\sigma(t,i)\ge\sigma_{\rm th}(t,i)$, $P_r$ reaches unity and the fibers break with probability $1$. Note also that this probabilistic model reduces to a conventional model in the limit of $T\rightarrow0$, in which a fiber can break only when $\sigma(i)\ge\sigma_{\rm th}(i)$.

The time evolution of the system is as follows. At $t=0$, the load is applied to the system, leading to a non-zero fracture probability of $P_r(0,i)$. This $P_r(0,i)$ is then compared with a random number $P^{\ast}(0,i)$ generated uniformly within the interval of $[0,1]$. If $P_r(0,i)>P^{\ast}(0,i)$, the $i$th fiber breaks and the borne load is redistributed. At the next time step ($t=1$), a different rupture probability is calculated based on the new local stress profile $\sigma(1, i)$. The model keeps evolving at each time by comparing the two probabilities, $P_r(t, i)$ and $P^{\ast}(t, i)$. The present authors adopt essentially the same time evolution rule in a fiber bundle model with a deterministic rupture rule (i.e., at $T=0$) and find creeplike time evolution \cite{royhatano1}. Here we extend this model to investigate the effect of stochasticity, the amplitude of which may be proportional to temperature.

Equation (\ref{Eq.1}) suggests that $P_r$ increases with temperature $T$ and the local stress $\sigma(t,i)$.  Then the nature of breakdown of the system will be determined by the interplay between the local stress profile, the individual strength of the fibers, and the temperature. In the present study, we concentrate on the mean field model, in which $\sigma(t,i)$ are uniform (independent of $i$) as a result of democratic load redistribution.

The stress threshold of each fiber,  $\sigma_{\rm th}(i)$, is a random quantity to be sampled from a certain probability distribution, $\rho(\sigma_{\rm th})$. Unless otherwise indicated, we adopt a uniform distribution with the mean of $1/2$ and the half-width of $\delta$:
\begin{equation}
\rho(\sigma_{\rm th}) = \begin{cases}
    1/2\delta,  & (1/2-\delta < \sigma_{\rm th} < 1/2 + \delta) \\
    0.  & ({\rm otherwise})
  \end{cases}
\end{equation}


\section{Numerical results}\label{Sec3}
Numerical results are produced for the mean field model with system sizes ranging between $10^3$ and $10^5$. To compute average values of physical quantities such as creep lifetime, $10^4$ to $10^5$ configurations are sampled at each system size. Here we have mainly studied the dispersion in creep lifetime and the evolution strain rate with time. Additionally, a series of dynamical transitions observed during the creep failure will be discussed in detail.
 

\subsection{Distribution of creep lifetime}
\begin{figure}[ht]
\centering
\includegraphics[width=8cm, keepaspectratio]{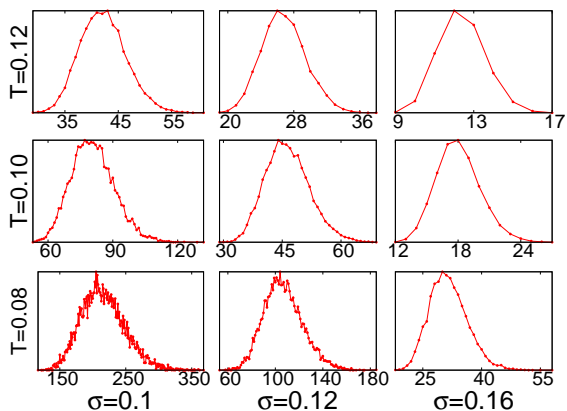} 
\caption{Distribution of creep lifetime for several parameter values of ($\sigma, T$). The average of the distribution shifts to lower value if either $\sigma$ or $T$ increases.} 
\label{Creep_Distribution}
\end{figure}
In this section, we discuss the distributions of creep lifetime in the parameter space of $\sigma-T$ to understand whether any extreme statistics is associated with it. Hereafter the creep lifetime is denoted as $\tau_c$, which varies from sample to sample. To compute the distributions of creep lifetime, $10^5$ samples are computed for the system size of $10^3$. 
Figure \ref{Creep_Distribution} shows that the peak of the distribution shifts to lower values as either of the parameters, temperature or applied stress, is increased. This happens as the probability of rupture increases with increasing $\sigma$ or $T$. A simultaneous increment in both parameters shifts the peak even faster.

\begin{figure*}[ht]
\centering
\includegraphics[width=14cm, keepaspectratio]{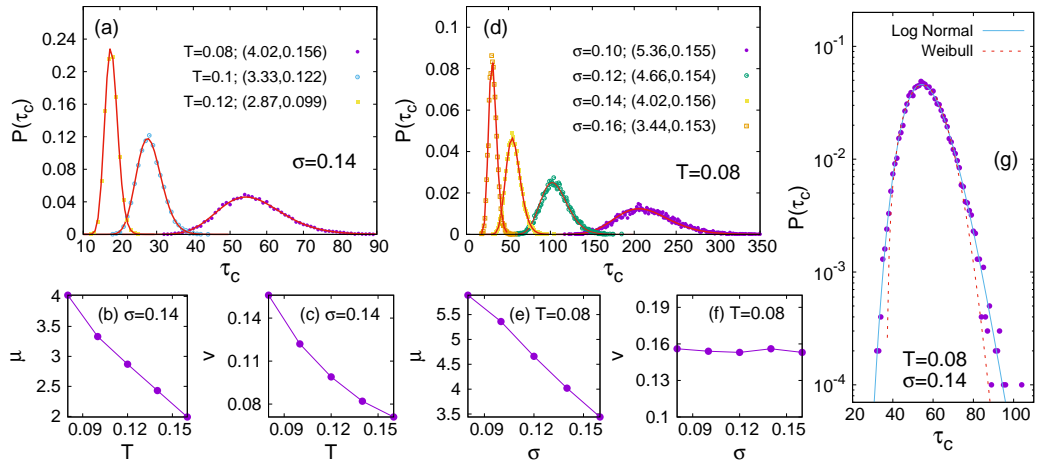} 
\caption{(a) Distribution of creep lifetime at applied stress $\sigma=0.14$ and temperature $T=0.08$, 0.1, and 0.12. The mean and the variance of $\log\tau_c$ depend on temperature $T$ as shown in (b) and (c), respectively. (d) Distribution of creep lifetime at temperature $T=0.08$ and applied stress $\sigma=0.1$, 0.12, 0.14, and 0.16. The mean and the variance of $\log\tau_c$ depend on the stress as shown in (e) and (f), respectively. (g) Lifetime distribution at $T=0.08$, $\sigma=0.14$ are fitted with log-normal distribution and three-parameter Weibull distribution. The comparison shows a better fit for log-normal distribution. $\delta$ is kept constant at 0.5 throughout the study.} 
\label{Creep_Distribution1}
\end{figure*}

The distribution of creep lifetime develops a tail, suggesting the existence of extreme statistics. As is shown in Fig.\ref{Creep_Distribution1}(a) and (b), a closer look on the distribution functions reveals that they are well fitted with the log-normal distribution given below.
\begin{equation}
P(\tau_c)=\displaystyle\frac{1}{\tau_cv\sqrt{2\pi}}\exp\left[-\displaystyle\frac{(\ln \tau_c-\mu)^2}{2v^2}\right],
\end{equation}  
where $v$ is the variance and $\mu$ is the mean of $\log\tau_c$. The values of $(\mu, v)$ are shown in the legends of individual panels of Fig. \ref{Creep_Distribution1}.
They depend on the applied stress and the temperature as shown in Fig.\ref{Creep_Distribution1} (b), (c), (e), and (f). We observe that the mean value decreases gradually as either the temperature or the applied stress increases. On the other hand, the variance is insensitive to the applied stress but decreases with increasing temperature. 

The lifetime distributions are also compared with the three-parameter Weibull distribution \cite{Dumonceaux}. The comparison is shown in Fig.\ref{Creep_Distribution1}(g) for $T=0.08$ and $\sigma=0.14$. Apparently, the log-normal distribution fits the simulation result better, particularly at the tails. This behavior remains at any temperature and stress as well as in the limit $T \rightarrow 0$, at which the rupture events are deterministic \cite{Royarxiv}.

\begin{figure}[ht]
\centering
\includegraphics[width=7cm, keepaspectratio]{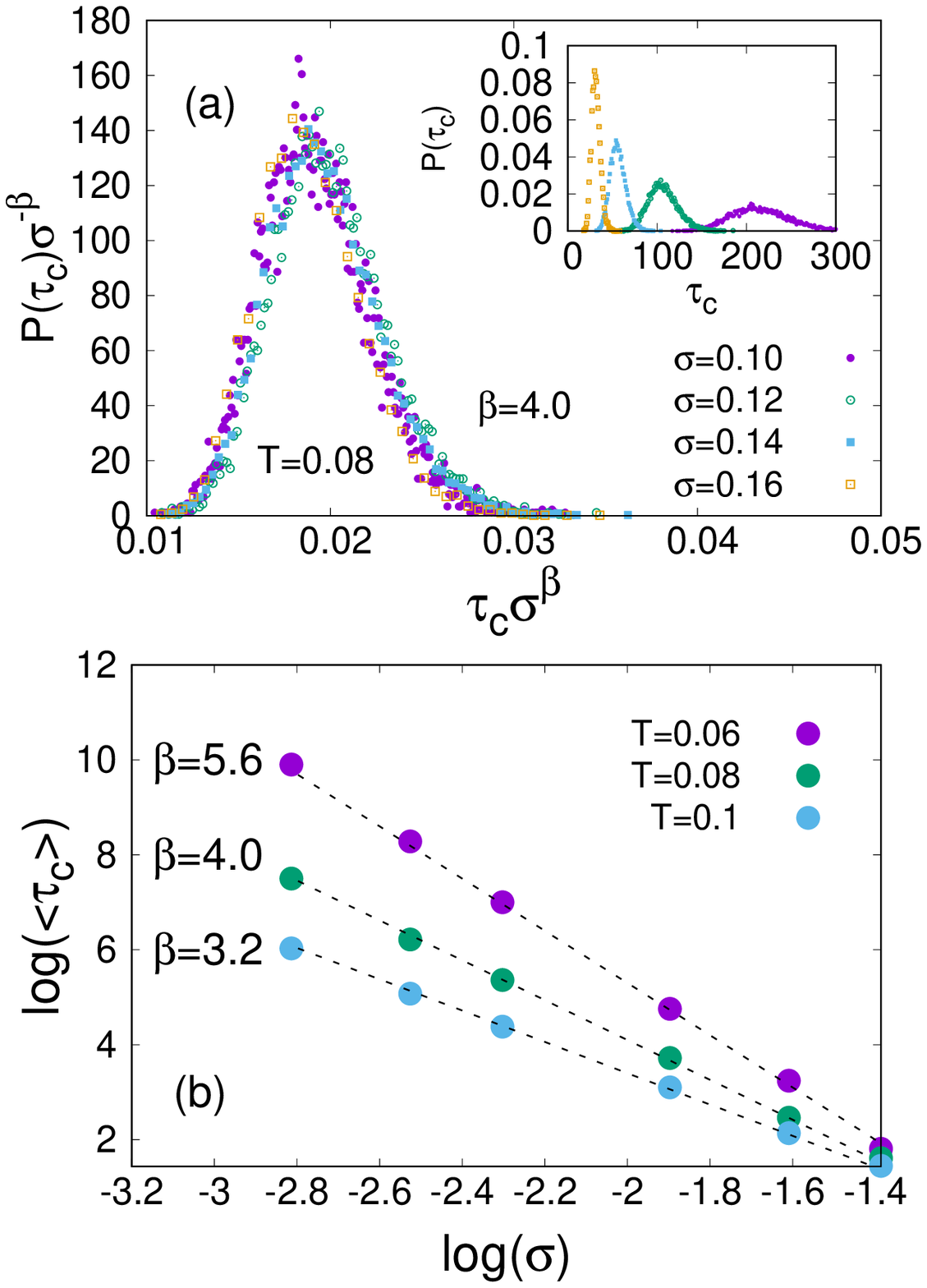} 
\caption{(a) At a constant $T$, as we vary the applied stress $\sigma$ we observe the following scaling: $\sigma^{-\beta}P(\tau_c) \sim \Phi(\tau_c \sigma^{\beta})$, where $\beta=4.0$. The inset shows the unscaled behavior. (b) Variation of average creep lifetime $\langle\tau_c\rangle$ with applied stress for three different temperatures. We keep $\delta$ fix at 0.5.} 
\label{Creep_Distribution2}
\end{figure}

At a constant temperature and varying applied stress, we observe the following scaling:
\begin{equation}
P(\tau_c) \simeq  \sigma^{\beta}\Phi(\tau_c \sigma^{\beta})
\end{equation}
with $\beta=4.0$. The scaling is demonstrated in Fig.\ref{Creep_Distribution2}(a) for different applied stress but at a constant temperature $T=0.08$. The inset of the same figure shows the unscaled behavior. The above scaling also tells that, at constant temperature $T$, the average  lifetime decreases with the applied stress in a scale-free manner with a temperature dependent exponent, $\beta$.
\begin{equation}\label{99}
\langle\tau_c\rangle \sim \sigma^{-\beta(T)} 
\end{equation}
The behavior is examined numerically and shown in Fig. \ref{Creep_Distribution2}(b). The exponent $\beta$ is a decreasing function of temperature. 


\subsection{Time evolution of strain rate} 
Next, we have studied the time evolution of the strain rate, $\dot{\epsilon}$.
Since the load $F$ is constant and the stiffness of the fibers are uniform, the strain ($\epsilon$) in the bundle is proportional to $F/L_t$, where $L_t$ denotes the number of intact fibers at time $t$. Then the strain rate is given by 
\begin{equation}
\dot{\epsilon}=\displaystyle\frac{d\epsilon}{dt}=\left(\displaystyle\frac{F}{L_{t+1}}-\displaystyle\frac{F}{L_{t}}\right),
\end{equation}
where the stiffness of the fibers are assumed to be unity.
\begin{figure}[ht]
\centering
\includegraphics[width=7cm, keepaspectratio]{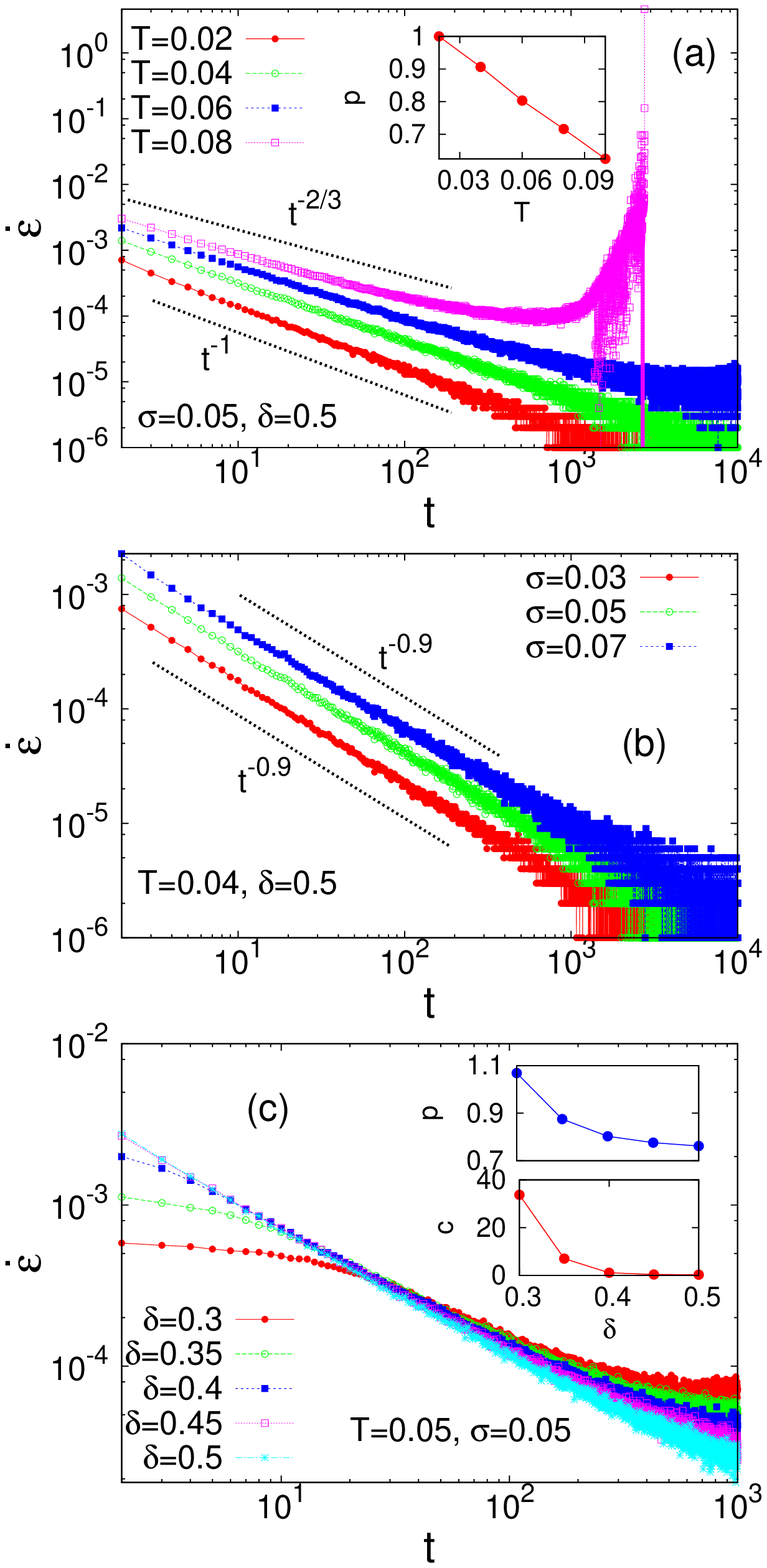}
\caption{(a) Time evolution of the strain rate $\dot{\epsilon}$ at several temperatures at $\sigma=0.05$ and $\delta=0.5$. The strain rate is fitted with $\dot{\epsilon}\sim t^{-p}$, where $p$ is a decreasing function of $T$ as shown in the inset. (b) Time evolution $\dot{\epsilon}$ for different applied stresses at $T=0.04$ and $\delta=0.5$. The exponent does not depend on $\sigma$. (c) Time evolution $\dot{\epsilon}$ for different strength of disorder at $T=0.05$ and $\sigma=0.05$. We observe $\dot{\epsilon}\sim (t+c)^{-p}$. Here both $c$ and $p$ are decreasing function of $\delta$ as shown in the inset.}
\label{Omori_Temp}
\end{figure}

Fig.\ref{Omori_Temp} shows the time evolution of strain rate for several sets of parameters: disorder $\delta$, applied stress $\sigma$ and temperature $T$. Fig.\ref{Omori_Temp}(a) shows the strain rate as a function of time at several values of temperature, where the stress and the disorder strength are common ($\sigma=0.05$ and $\delta=0.5$). We observe the following power-law behavior for the strain rate $\dot{\epsilon}$.
\begin{equation}
\label{omori}
\dot{\epsilon} \sim \displaystyle\frac{1}{(t+c)^p},
\end{equation} 
where $t$ is the time elapsed after the loading and $c$ is the time constant for the onset of scale-free behavior. This is essentially identical to Andrade creep, $t^{-p}$, if the time constant $c$ is negligible. However, note that the time constant $c$ is nonzero in the present model as shown in Fig.\ref{Omori_Temp}(c). The time constant $c$ increases as the extent of disorder $\delta$ decreases  as shown in the inset of Fig.\ref{Omori_Temp}(c), but insensitive to the applied stress. 

The exponent $p$ depends on temperature, and decreases gradually as the temperature increases as shown in Fig.\ref{Omori_Temp}(a). The inset shows the variation of $p$ with respect to the temperature. In a wide temperature range ($0.02\le T\le  0.1$), the exponent $p$ changes gradually from $1$ to $0.67$. This range of exponent values is in good agreement with experiments. The exponent $p$ is insensitive to the applied stress $\sigma$ as shown in Fig.\ref{Omori_Temp}(b), and depends only gradually on the dispersion in the threshold distribution as shown in the inset of Fig.\ref{Omori_Temp}(c).

\begin{figure}[ht]
\centering
\includegraphics[width=7cm, keepaspectratio]{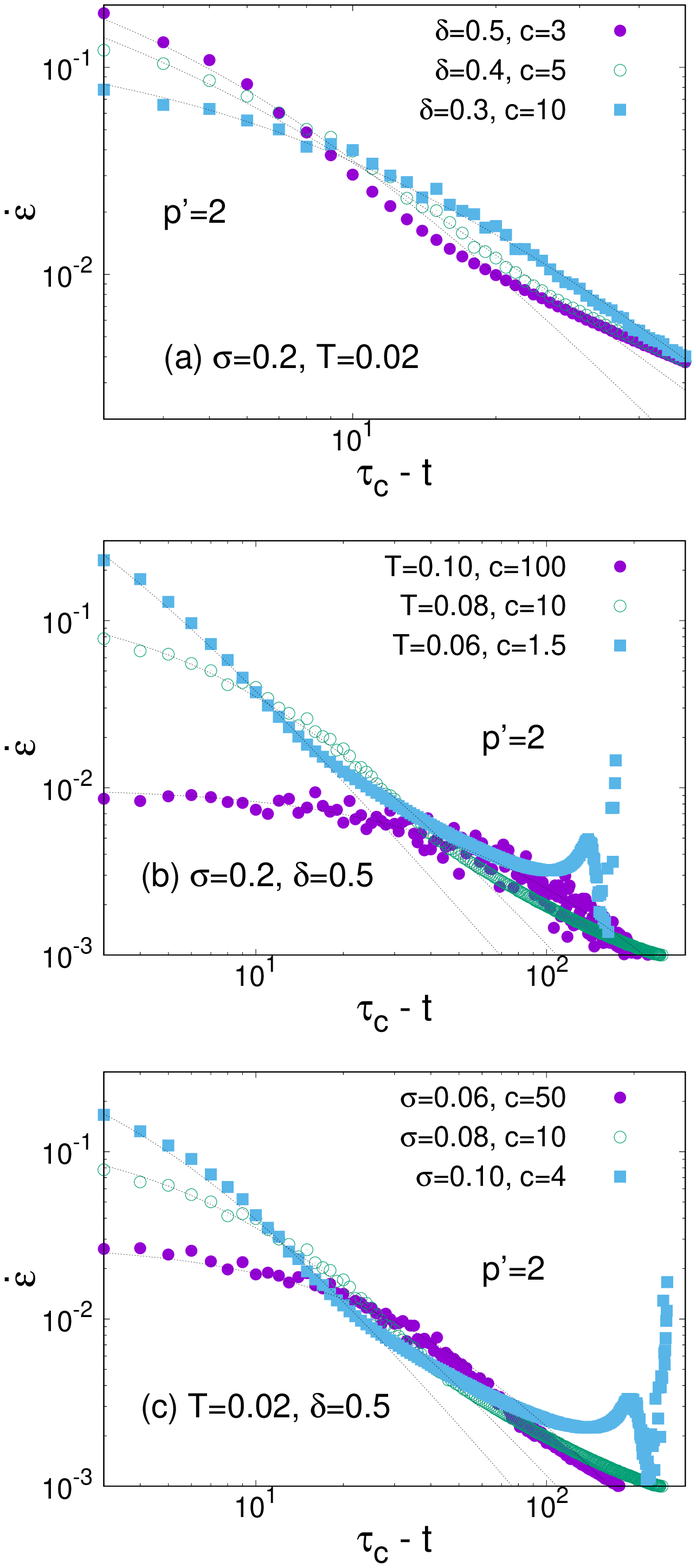} 
\caption{Strain rate as a function of $\tau_c-t$. The dotted lines are the inverse Omori law: $\dot{\epsilon}\sim [(\tau_c-t)+c]^{-p^{\prime}}$. The exponent $p^{\prime}=2$ appears to be robust in all the panels. (a) The time constant $c$ decreases as $\delta$ increases. (b) and (c) show that the $c$-value decreases as $T \rightarrow 0$ and $\sigma \rightarrow \sigma_c$. } 
\label{Inv_Omori}
\end{figure}

In the tertiary creep, in which deformation accelerates toward breakdown, the strain rate can be fitted with a power law again as shown in Fig. \ref{Omori_Temp}:
\begin{equation}
\label{inv_omori}
\dot{\epsilon}\sim \frac{1}{[(\tau_c-t)+c]^{p^{\prime}}},
\end{equation}
which is known as the inverse Omori law. Importantly, the characteristic time $c$ is nonzero again. As shown in Fig. \ref{Inv_Omori} (a), the $c$-value increases as the extent of disorder $\delta$ declines. Figure \ref{Inv_Omori} (b) and (c) show that the $c$-value also increases as the temperature increases (b), or the stress decreases (c). Namely, the acceleration lasts longer towards breakdown at lower temperature, lower stress, and/or stronger disorder. In terms of predictability of the creep lifetime, the number of parameters should be as small as possible. In this respect, the failure may be more predictable for vanishing $c$ (i.e., large disorder). This tendency of predictability is consistent with an experiment with controllable disorder \cite{Vasseur}. On the other hand, the exponent $p{\prime}$ appears to remain for the parameter range investigated here, although the power law itself is less clear if $c$ is large. 

The above two power-law behaviors, the Omori and the inverse Omori laws, are common to a simple deterministic model at $T=0$ \cite{royhatano1}. However, the parameters behave differently in most respects. For $T=0$ case, the exponent $p^{\prime}$ is approximately $2$ irrespective of the randomness and the stress, whereas it is a decreasing function of disorder and temperature in the present case. For the primary creep, the characteristic time $c$ in Eq. (\ref{omori}) is insensitive to the stress in the present model, whereas it increases with the stress at $T=0$. For the tertiary creep, the characteristic time $c$ in Eq. (\ref{inv_omori}) vanishes at $T=0$, whereas it is nonzero in the present model. Contrastingly, the temperature dependence of the characteristic time $c$ in Eq. (\ref{inv_omori}) extrapolates to the zero temperature case.
 

\subsection{Transitions in abruptness} 
In this section, we have studied the abruptness of failure. Denoting the number of unbroken fibers as $L_t$, we define the decreasing rate of fibers as $s(t)=L_{t}-L_{t+1}$. Then the time average of $s(t)$ is computed for each sample from $t=0$ to the time of creep failure, $\tau_c$. This time-averaged value of $s(t)$ is denoted by $s$. For instance, $s=L$ if $\tau_c =1$, and $s=L/2$ if $\tau_c=2$, and so forth. (Here $L=L_0$.) Indeed, one can easily confirm that $s=L/\tau_c$. Larger $s$ thus implies that the breakdown is more abrupt. Since this $s$ varies from sample to sample, the ensemble average, which is denoted by $\langle s \rangle$, is taken over $10^4$ configurations. Note that $\langle s \rangle/L = \langle\tau_c^{-1}\rangle$.

\begin{figure}[ht]
\centering
\includegraphics[width=8cm, keepaspectratio]{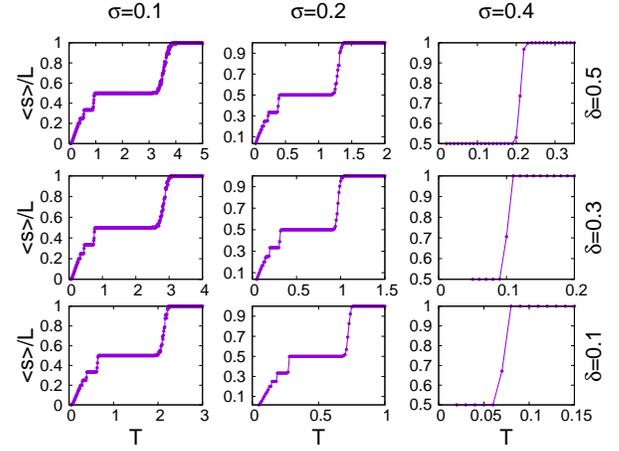} 
\caption{$\langle s \rangle/L$ v/s $T$ for different temperature (horizontally) and external stress (vertically) values. The average avalanche values decrease in jumps when $T$ is decreased. Number of such jumps as well as the position of such jumps are affected by $\sigma$, while $\delta$ only affects the position of jumps.}  
\label{Dynamical_Sigma_Delta}
\end{figure}

Figure \ref{Dynamical_Sigma_Delta} shows the behavior of $\langle s \rangle/L$. The applied stress $\sigma$ is varied from 0.1 to 0.4, while the disorder strength $\delta$ ranges in between 0.1 and 0.3. At high temperatures, $\langle s \rangle \sim L$, suggesting that the total bundle breaks in a single time step. As $T$ decreases, $\langle s \rangle/L$ decreases abruptly at a certain temperature. As the temperature decreases further, similar drops of $\langle s \rangle/L$ occur repeatedly but the amplitude of abrupt change declines gradually at lower temperatures.
The jump heights gradually decreases as we go to lower $T$ values. The first jump is from $1.0$ to $0.5$, and the second jump is from $0.5$ to $1/3$, and so on. The height of $n$th jump, denoted by $\Delta h_n$, is  
\begin{equation}
\Delta h_n/L \simeq \displaystyle\frac{1}{n(n+1)}.
\end{equation}
This is understandable since $\langle s \rangle/L = \langle \tau_c\rangle$ and $\tau_c$ goes from $n$ to $n+1$ for the $n$th jump. Figure \ref{Dynamical_Sigma_Delta} also shows that the jumps gradually disappear as the stress $\sigma$ increases. For example, there are a number of jumps with decreasing jump heights for $\sigma=0.1$. On the other hand, for $\sigma=0.4$, we observe only one jump from $\langle s \rangle/L=1$ to 0.5. This is because the failure occurs very abruptly at higher stresses, taking only a few time steps. Thus, there are only a few large avalanches at higher stress, and subsequent transitions to smaller avalanches are absent. This behavior remains almost unaltered with increasing disorder strength, whereas the transition temperatures are affected. 
 
\begin{figure}[t]
\centering
\includegraphics[width=8cm, keepaspectratio]{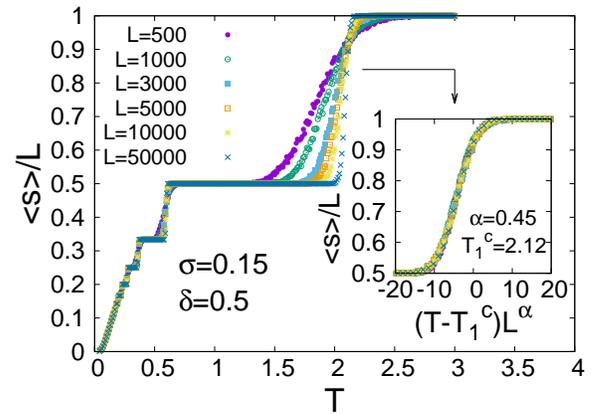} 
\caption{Variation of average avalanche size $\langle s \rangle/L$ with increasing temperature $T$ at applied stress $\sigma=0.15$ and disorder strength $\delta=0.5$. The inset shows the system size scaling where $\langle s \rangle/L$ deviates from unity: $\langle s \rangle/L \sim \Psi[(T-T^c_1)L^{\alpha}]$ where $\alpha=0.45$ and $T^c_1=2.12$.}  
\label{Avalanche_Size}
\end{figure}

\begin{figure}[ht]
\centering
\includegraphics[width=7cm, keepaspectratio]{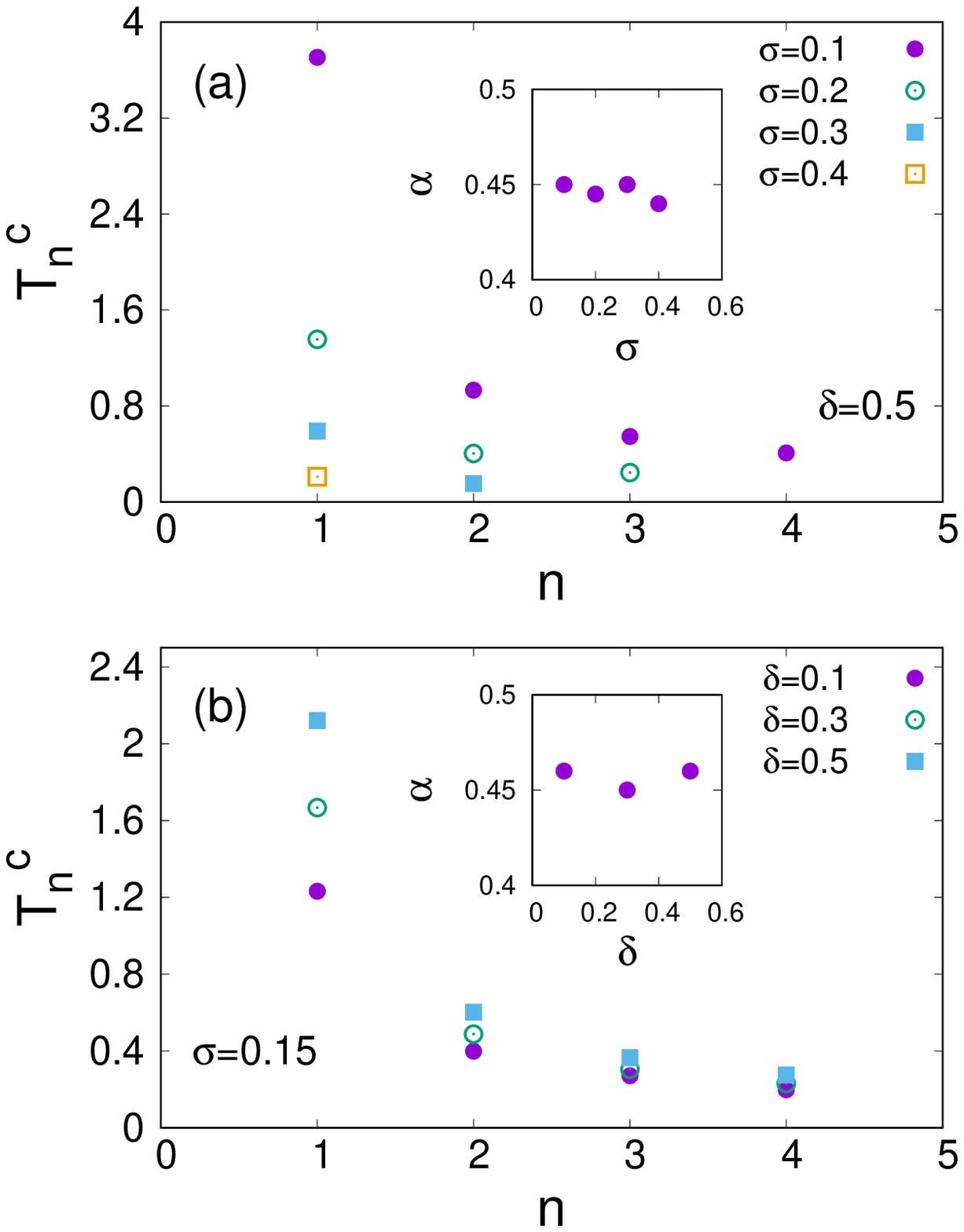} 
\caption{Variation of the critical temperature $T_n^c$ with the number $n$ of the jumps for (a) different $\sigma$ at constant $\delta$ (=0.5) as well as (b) for different $\delta$ and constant $\sigma$ (=0.15).}  
\label{Critical_Temp}
\end{figure}

To study these jumps more closely, we have carried out the finite size scaling. Figure \ref{Avalanche_Size} shows the behavior of $\langle s \rangle/L$ at several system sizes ranging from $L=5\times10^2$ to $5\times10^4$. The system size scaling is given as follows: 
\begin{equation}
\label{Eq.n1}
\langle s \rangle/L \sim \Psi\big[(T-T_n^c)L^{\alpha}\big],
\end{equation}
where $T_n^c$ is the transition temperature for the $n$th jump, and the exponent $\alpha$ is a constant. The inset of Fig.\ref{Avalanche_Size} shows the system size scaling for $n=1$ with $T_1^c=2.12$. The exponent $\alpha$ is estimated as $0.45$ irrespective of the stress and the temperature. We get a series of $T_n^c$ values from the same scaling for different jumps. This scaling relation suggests that jumps in $\langle s \rangle/L$ with varying temperature are indeed phase transitions.

\begin{figure}[ht]
\centering
\includegraphics[width=9cm, keepaspectratio]{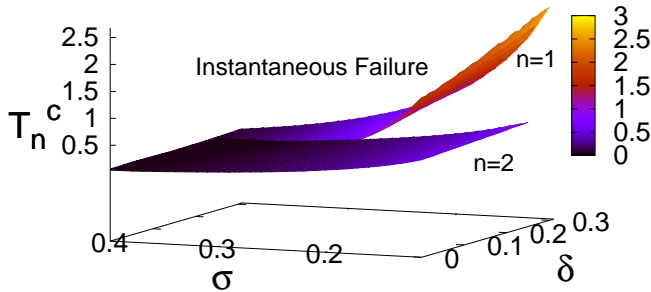} 
\caption{Variation of the critical temperature $T_n^c$ with a continuous variation in applied stress $\sigma$ and strength of disorder $\delta$. The variation is shown for: (a) the first transition (n=1) and (b) the second one (n=2).}  
\label{Phase_Diagram2}
\end{figure}

Fig.\ref{Critical_Temp}(a) and Fig.\ref{Critical_Temp}(b) shows the variation of $T_n^c$ with $n$ for different values of $\sigma$ and $\delta$. Both behaviors suggest a decrease in $T_n^c$ with $n$. In addition, the  $n$ values itself decreases as we go to higher $\sigma$ values, suggesting a lesser number of jumps and at the same time big avalanches. On the other hand, $n$ values remain invariant of the disorder strength. For a particular jump (n-value), the critical temperature decreases with increasing applied stress and decreasing strength of disorder.                

Finally, Fig.\ref{Phase_Diagram2} shows the variation of $T_n^c$ for the first and second transitions, when both applied stress $\sigma$ and strength of disorder $\delta$ varies simultaneously. The figure shows two planes corresponding to the first ($n=1$) and second ($n=2$) transitions producing three separate regions. The region above the plane $n=1$ corresponds to instantaneous failure where the total bundle breaks in a single avalanche producing a unit creep lifetime ($\tau_c=1$). In this region, the distribution of $\tau_c$ is a delta function at 1.   



\section{Discussions}\label{Sec4}
In the present work, the effect of temperature in creep failure is probed by a fiber bundle model with a probabilistic algorithm. As a result of the stochasticity, the model can fail even if the applied stress is below the critical value at $T=0$. The dynamical properties of the model, including the exponents for power-law behaviors, are comparable to those of real materials. The model exhibits a large scatter in the creep lifetime, which can be described by a log-normal distribution with the temperature-dependent mean and variance. The mean lifetime depends on the applied stress in a scale-free manner.  A detailed study of the time-dependent strain rate enables us to detect the characteristic power laws with a nonzero time constant. The exponents depends slightly on the temperature, the range is comparable with the phenomenology known as the Andrade creep.

For a practical purpose, the average lifetime alone does not provide us with sufficient information on creep failure since the lifetime distribution is skewed \cite{Xing}. Both the log-normal and the Weibull distributions are observed for the lifetime of some alloys \cite{Evans,Evans2,Evans1}. The Weibull distribution was also found for STS304 stainless steel \cite{Kim}. In our study, the creep lifetime distribution is fitted more convincingly with the log normal distribution than the Weibull distribution. Another major difference is in our case the Weibull distribution contains three parameters while the above experimental creep lifetimes are fitted well with two-parameter Weibull distribution. In addition, the stress dependence of the lifetime in our model agrees well with experimental behaviors \cite{Proceeding,Evans2,Benaarbia}. Overall, the present numerical results are mostly consistent with experimental results on materials 
\cite{Proceeding,Evans,Evans1,Evans2,Benaarbia,Jagla,Fabeny,Lahyani,Xing,Kim}. 

Understanding of creep behavior is a fundamental problem in material science and engineering. Prior knowledge on the creep lifetime for a particular material reduces the threat of sudden catastrophic failure and increases predictability in the failure process. At the same time, sufficient knowledge of creep dynamics makes us aware if the system is approaching the global failure. In this paper, we have presented a detailed study of the creep dynamics in a disordered system. 
Our next step is to explore the model with other modifications such as the inclusion of stress concentration around a broken fiber. Application of time-dependent loading allows us to explore the possibility of fatigue within the model. 
 



\end{document}